\documentclass[twocolumn,superscriptaddress,pra,aps,10pt,nofootinbib]{revtex4-1}
\usepackage{times}
\usepackage{amssymb}
\usepackage{color}
\usepackage{amsmath}
\usepackage{mathrsfs}
\usepackage{amsbsy}
\usepackage{amsthm}
\usepackage{graphicx}
\usepackage{bbm}
\usepackage{bm}
\usepackage{epsfig}
\usepackage{xfrac}
\usepackage{xcolor}

\definecolor{myurlcolor}{rgb}{0,0,0.7}
\definecolor{myrefcolor}{rgb}{0.8,0,0}
\usepackage[unicode=true,pdfusetitle, bookmarks=true,bookmarksnumbered=false,bookmarksopen=false, breaklinks=false,pdfborder={0 0 0},backref=false,colorlinks=true, linkcolor=myrefcolor,citecolor=myurlcolor,urlcolor=myurlcolor]{hyperref}

\usepackage{braket}
\renewcommand{\v}[1]{\ensuremath{\mathbf{#1}}} 
\newcommand{\gv}[1]{\ensuremath{\text{\boldmath$ #1 $}}}
\newcommand{\abs}[1]{\left| #1 \right|} 
\newcommand{\norm}[1]{\left\| #1 \right\|} 
\newcommand{\trace}{\mathrm{Tr}}
\newcommand{\tl}{{\textsc{l}}}

\newcommand{\ts}{{\textsc{s}}}
\newcommand{\mS}{{\mathcal{S}}}

\newcommand{\mH}{{\mathcal{H}}}

\newcommand{\mR}{{\mathcal{R}}}
\newcommand{\frakF}{{\mathfrak{F}}}
\newcommand{\tLambda}{{\tilde{\Lambda}}}
\newcommand{\tJ}{{\tilde{J}}}
\newcommand{\mP}{{\mathcal{P}}}

\newcommand{\mE}{{\mathcal{E}}}

\newcommand{\id}{{\mathbbm{1}}}
\newcommand{\balpha}{\bar{\alpha}}

\newcommand{\frakh}{{\mathfrak{h}}}
\newcommand{\dfrakh}{{\Delta\mathfrak{h}}}

\newcommand{\frakj}{{\mathfrak{j}}}

\newcommand{\scrJ}{{\mathscr{J}}}
\newcommand{\scrA}{{\mathscr{A}}}

\newcommand{\vh}{{\v{h}}}
\newcommand{\inull}{{\mathfrak{n}}}
\newcommand{\is}{{\mathfrak{s}}}
\newcommand{\iw}{{\bar{\mathfrak{s}}}}

\newcommand{\dvh}{{\Delta \v{h}}}

\newcommand{\vj}{{\gv{j}}}

\newcommand{\vv}{{\gv{v}}}
\newcommand{\vw}{{\gv{w}}}

\newcommand{\vu}{{\gv{u}}}
\newcommand{\vL}{{\v{L}}}

\newcommand{\vJ}{{\v{J}}}

\newcommand{\bR}{{\mathbb{R}}}
\newcommand{\bC}{{\mathbb{C}}}
\renewcommand{\Re}{{\mathrm{Re}}}

\renewcommand{\epsilon}{\varepsilon}
\newcommand{\appropto}{\mathrel{\vcenter{
  \offinterlineskip\halign{\hfil$##$\cr
    \propto\cr\noalign{\kern2pt}\sim\cr\noalign{\kern-2pt}}}}}
\newcommand{\dket}[1]{\vert {#1} \rangle \! \rangle} 
\newcommand{\dbra}[1]{\langle \! \langle {#1} \vert} 
\newcommand{\dbraket}[1]{\langle \! \langle {#1} \rangle \! \rangle}
\let\baraccent=\= 
\renewcommand{\=}[1]{\stackrel{#1}{=}} 

\newcommand{\thmref}[1]{\hyperref[#1]{Theorem~\ref{#1}}}
\newcommand{\lemmaref}[1]{\hyperref[#1]{Lemma~\ref{#1}}}
\newcommand{\figref}[1]{\hyperref[#1]{Fig.~\ref{#1}}}
\newcommand{\figaref}[1]{\hyperref[#1]{Fig.~\ref{#1}a}}
\newcommand{\figbref}[1]{\hyperref[#1]{Fig.~\ref{#1}b}}
\newcommand{\figcref}[1]{\hyperref[#1]{Fig.~\ref{#1}c}}
\renewcommand{\eqref}[1]{\hyperref[#1]{Eq.~(\ref{#1})}}
\newcommand{\eqsref}[2]{\hyperref[#1]{Eqs.~(\ref{#1})-(\ref{#2})}}
\newcommand{\appref}[1]{\hyperref[#1]{Appx.~\ref{#1}} of \cite{SM}}

\usepackage{dsfont}

\newtheorem{lemma}{Lemma}

\begin{document}

\title{
Optimal Approximate Quantum Error Correction for Quantum Metrology
}

\author{Sisi Zhou}
\affiliation{Departments of Applied Physics and Physics, Yale University, New Haven, Connecticut 06511, USA}
\affiliation{Yale Quantum Institute, Yale University, New Haven, Connecticut 06511, USA}

\author{Liang Jiang}
\affiliation{Departments of Applied Physics and Physics, Yale University, New Haven, Connecticut 06511, USA}
\affiliation{Yale Quantum Institute, Yale University, New Haven, Connecticut 06511, USA}
\affiliation{Pritzker School of Molecular Engineering, The University of Chicago, Illinois 60637, USA}

\date{\today}

\begin{abstract}
For a generic set of Markovian noise models, the estimation precision of a parameter associated with the Hamiltonian is limited by the $1/\sqrt{t}$ scaling where $t$ is the total probing time, in which case the maximal possible quantum improvement in the asymptotic limit of large $t$ is restricted to a constant factor. However, situations arise where the constant factor improvement could be significant, yet no effective quantum strategies are known. Here we propose an optimal approximate quantum error correction (AQEC) strategy asymptotically saturating the precision lower bound in the most general adaptive parameter estimation scheme where arbitrary and frequent quantum controls are allowed. We also provide an efficient numerical algorithm finding the optimal code. Finally, we consider highly-biased noise and show that using the optimal AQEC strategy, strong noises are fully corrected, while the estimation precision depends only on the strength of weak noises in the limiting case. 
\end{abstract}

\maketitle

\paragraph*{Introduction.--}

Quantum metrology is one of the most important state-of-the-art quantum technologies, studying the precision limit of parameter estimation in quantum systems~\cite{giovannetti2006quantum,giovannetti2011advances,degen2017quantum,braun2018quantum,pezze2018quantum,pirandola2018advances}. The task involves preparing a suitable initial state of the system, allowing it to evolve under quantum controls for a specific time, performing a suitable measurement, and inferring the value of the unknown system parameter from the measurement outcome. To enhance the estimation precision, a variety of quantum strategies have been proposed, such as squeezing the initial state~\cite{caves1981quantum,wineland1992spin,kitagawa1993squeezed,huelga1997improvement,ulam2001spin,demkowicz2013fundamental}, optimizing the probing time~\cite{chaves2013noisy}, monitoring the environment~\cite{plenio2016sensing,albarelli2017ultimate,albarelli2019restoring}, exploiting non-Markovian effects~\cite{matsuzaki2011magnetic,chin2012quantum,smirne2016ultimate}, optimizing the control Hamiltonian~\cite{yuan2016sequential,liu2017quantum,xu2019transferable} and quantum error correction~\cite{kessler2014quantum,arrad2014increasing,dur2014improved,ozeri2013heisenberg,reiter2017dissipative,lu2015robust,sekatski2017quantum,demkowicz2017adaptive,zhou2018achieving,layden2018spatial,layden2019ancilla,gorecki2019quantum}.

Quantum mechanics places a fundamental limit on estimation precision, the Heisenberg limit (HL), where the estimation precision scales like $1/N$ for $N$ probes; or equivalently, $1/t$ for a total probing time $t$. In the noiseless case, the HL is achievable using the maximally entangled state among probes~\cite{leibfried2004toward,giovannetti2006quantum}. In practice, decoherence plays an indispensible role. Under many typical noise models, the estimation precision will follow the standard quantum limit (SQL) with scaling $1/\sqrt{N}$ (or $1/\sqrt{t}$)~\cite{fujiwara2008fibre,escher2011general,ji2008parameter,demkowicz2012elusive,demkowicz2014using,kolodynski2013efficient,sekatski2017quantum,demkowicz2017adaptive,zhou2018achieving}, the same as the central limit theorem scaling using classical strategies. Nevertheless, the superiority of quantum strategies over classical strategies by a constant-factor improvement, as opposed to a scaling improvement, was proven in several cases~\cite{ulam2001spin,escher2011general,demkowicz2014using}. There were also situations where the HL is achievable using quantum strategies even in the presence of noise~\cite{zhou2018achieving,albarelli2019restoring}. 

Due to the difficulty in obtaining the exact precision limits for general noise models using different quantum strategies, several asymptotical lower bounds have been proposed~\cite{fujiwara2008fibre,ji2008parameter,escher2011general,demkowicz2012elusive,kolodynski2013efficient,demkowicz2014using,sekatski2017quantum,demkowicz2017adaptive,zhou2018achieving,czajkowski2019many}. For example, the channel simulation method was used to prove the SQL lower bound for programmable channels~\cite{ji2008parameter,demkowicz2012elusive,demkowicz2014using}. A necessary and sufficient condition of achieving the HL under Markovian noise was established using the channel extension method~\cite{sekatski2017quantum,demkowicz2017adaptive,zhou2018achieving}. Although these bounds have been successful at showing the scaling limit of quantum strategies, only in several special cases, the saturability of these lower bounds was established, e.g. for dephasing and erasure noise~\cite{demkowicz2014using} and for teleportation-covariant channels as a special type of programmable channels~\cite{pirandola2017ultimate,laurenza2018channel}. 
A saturability statement of the SQL lower bound under general noise models and an efficient algorithm solving the optimal strategy remain missing up to the present day.  

\begin{figure}[ht]
\includegraphics[width=7cm]{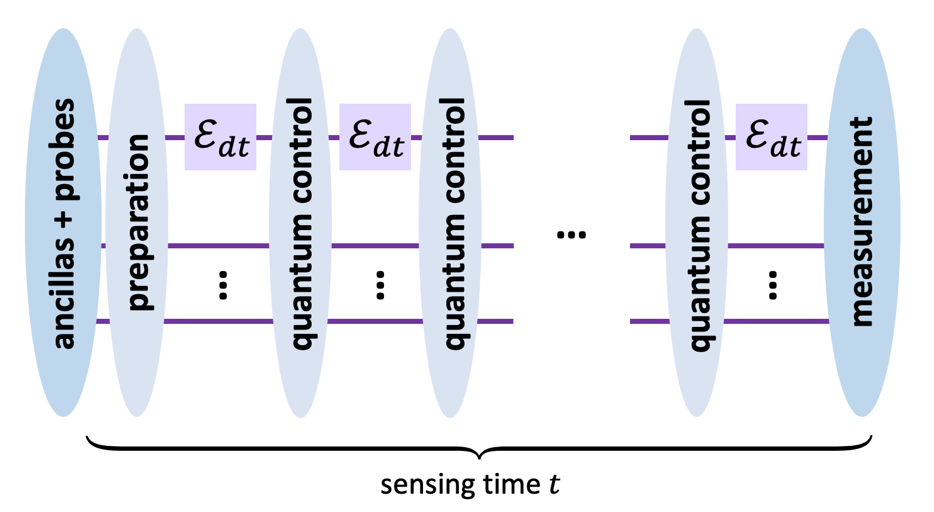}
\caption{\label{fig:strategy} The most general adaptive sequential strategy
where one probe sequentially senses the parameter for time $t$, with quantum controls (arbitrary completely positive and trace-preserving (CPTP) maps) applied every $dt$ and an arbitrary number of noiseless ancillas available. 
$\mE_{dt}(\rho) = \rho + ( - i[\omega H,\rho] + \sum_{i=1}^r L_i \rho L_i^\dagger - \frac{1}{2}\{L_i^\dagger L_i,\rho\} ) dt + O(dt^2)$ describes the evolution of the probe in an infinitesimally small time interval $dt$. 
}
\end{figure}

We address both of these open questions in this work. Here we consider parameter estimation under general Markovian noise using the most general adaptive sequential strategy (see \figref{fig:strategy}). We propose an approximate quantum error correction (AQEC) strategy saturating the SQL lower bound of precision (asymptotically) and an efficient numerical algorithm solving the optimal AQEC codes for different noises. The saturability of the precision lower bound we prove here not only answers an important question in quantum metrology theory, but also paves the way for identifying the optimal quantum strategies in future experiments. 

Quantum error correction (QEC)~\cite{nielsen2002quantum} was first shown useful in quantum metrology in a typical scenario where the dephasing noise in a qubit probe is corrected by QEC, while the $X$ (the Pauli-X operator) signal remains intact~\cite{kessler2014quantum,arrad2014increasing,dur2014improved,ozeri2013heisenberg}. Later on, the result was generalized to arbitrary Markovian noise~\cite{demkowicz2017adaptive,zhou2018achieving}, stating that the HL is achievable using the sequential QEC strategy if and only if the HNLS condition is satisfied, i.e. the signal Hamiltonian is not in the Lindblad span---an operator subspace defined using Lindblad operators~\cite{gorini1976completely,lindblad1976generators,breuer2002theory}. In practice, however, HNLS is often violated and the estimation precision is limited by the SQL, for example, sensing any single-qubit signal under depolarizing noise. Standard QEC strategies would be useless in this case, as the signal will be completely eliminated if the noise is fully corrected. However, here we show that by performing the QEC in an approximate fashion, the highest possible precision limit is achievable, marking another triumph of the QEC strategy in quantum metrology. 

In this Letter, we first review the SQL precision lower bound under Markovian noise using the sequential strategy when HNLS is violated. Then we describe our AQEC strategy consisting of both a two-dimensional AQEC code and an optimal recovery channel. This allows the original quantum channel to be reduced to an effective channel where a $Z$ (the Pauli-Z operator) signal was sensed under dephasing noise---a special case where the precision lower bound was known to be saturable~\cite{ulam2001spin,escher2011general,demkowicz2014using}. Finally, we optimize the achievable precision over all possible AQEC codes, which coincides with the precision lower bound, completing the proof.

\paragraph*{Precision lower bound.--} We assume the evolution of the quantum system is described by the following quantum master equation~\cite{gorini1976completely,lindblad1976generators,breuer2002theory}:
\begin{equation}
\label{eq:master}
\frac{d\rho}{dt} = -i[\omega H,\rho] + \sum_{i=1}^r \Big(L_i \rho L_i^\dagger - \frac{1}{2}\{L_i^\dagger L_i,\rho\}\Big),
\end{equation}
where $\omega$ is the unknown parameter, $\rho\in \mH_S \otimes \mH_A$, $\mH_S$ is the probe space $H$ and $\{L_i\}_{i=1}^r$ acting on ($H$, $L_i$ are shorthand for $H\otimes \id$, $L_i \otimes \id$, respectively) and $\mH_A$ is the noiseless ancillary space (see \figref{fig:strategy}). We assume $I,\{L_i\}_{i=1}^r$  are linearly independent, $\dim \mH_S = d$ and $\dim \mH_A = 2d$. The Lindblad span associated with \eqref{eq:master} is $\mS = {\rm span}\{\id,L_i,L_i^\dagger,L_i^\dagger L_j,\forall i,j\}$, where ${\rm span}\{\cdot\}$ denotes the real linear subspace of Hermitian operators spanned by $\{\cdot\}$. According to the quantum Cram\'{e}r-Rao bound~\cite{helstrom1968minimum,helstrom1976quantum,braunstein1994statistical,paris2009quantum}, the standard deviation $\delta \omega$ of the $\omega$-estimator is bounded by $\delta\omega \geq (N_{\rm expr} F(t))^{-1/2}$, where $N_{\rm expr}$ is the number of experiments and $F(t)$ is the so-called quantum Fisher information (QFI) as a function of the final state $\rho(t)$. The bound is asymptotically saturable using the maximum likelihood estimator as $N_{\rm expr}$ goes to infinity~\cite{casella2002statistical,lehmann2006theory}. Therefore, 
finding the optimal sequential strategy boils down to maximizing $F(t)$ over all input states and quantum controls. For an input state $\ket{\psi}$ evolving noiselessly under Hamiltonian $\omega H$, $F(t) = 4t^2 (\bra{\psi}H^2\ket{\psi} - (\bra{\psi}H\ket{\psi})^2)$ and $\delta \omega \propto 1/t$ follows the HL. In the noisy case, it was proven that the HL is achievable if and only if $H \notin \mS$ (the HNLS condition) and there exists a QEC strategy achieving the HL~\cite{demkowicz2017adaptive,zhou2018achieving}.   

{The HNLS condition holds usually when the noise has a special structure, e.g. rank-one noise~\cite{sekatski2017quantum} or spacially correlated noise~\cite{layden2018spatial,layden2019ancilla}. 
For generic noise, however, the HNLS condition is often violated. In this Letter, we focus on the latter situation where $H \in \mS$ and the QFI follows the SQL}~\cite{demkowicz2017adaptive,zhou2018achieving}:
\begin{equation}
\label{eq:upper}
F(t) \leq 4t \min_{h,\vh,\frakh|\beta = 0}\norm{\alpha},
\end{equation}
where $\norm{\cdot}$ is the operator norm of a matrix, $h\in\bR$, $\vh \in \bC^r$, $\frakh \in \bC^{r\times r}$ is hermitian, 
\begin{align}
\label{eq:alpha}
\alpha &= (\vh\id + \frakh \vL)^\dagger (\vh\id + \frakh \vL)
,
\\
\label{eq:beta}
\beta &= H + h\id + \vh^\dagger \vL + \vL^\dagger \vh + \vL^\dagger \frakh \vL
,
\end{align}
where $\vL := (L_1,L_2,\ldots,L_r)^T$ and $\vh\id := (\vh_1\id,\ldots,\vh_r\id)^T$. 
Here we introduce an AQEC strategy which (asymptotically) saturates the QFI upper bound up to an arbitrarily small error under arbitrary Markovian noise. That is, for any small $\delta > 0$, there exists an AQEC strategy such that 
\begin{equation}
\frakF:=\max_{t > 0}\frac{F(t)}{t} = 4\min_{h,\vh,\frakh|\beta = 0}\norm{\alpha} - \delta,
\end{equation}
where we define the normalized QFI $\frakF$ as the objective function we maximize. The upper bound is saturated asymptotically in the sense that $\lim_{t\rightarrow \infty} F(t)/t = \frakF$.

\paragraph*{Approximate quantum error correction.--}

Here we propose a set of AQEC codes for quantum metrology and show that the effective channel under fast AQEC is  {an effective qubit dephasing channel in the logical space}. In this way, identifying the optimal recovery channel for quantum metrology is equivalent to minimizing the noise rate of the dephasing channel where a closed-form solution exists, as opposed to generic AQEC scenarios where many known AQEC recovery channels are only suboptimal~\cite{barnum2002reversing,fletcher2007optimum,beny2010general,ng2010simple,tyson2010two,albert2018performance}.

Let $P$ be the projection on to the code space $\ket{0_\tl}\bra{0_\tl} + \ket{1_\tl}\bra{1_\tl}$, where $\ket{0_\tl}$ and $\ket{1_\tl}$ are the logical zero and one states. Applying the AQEC quantum operation $\mP + \mR \circ \mP_\perp$ infinitely fast, the effective evolution would be (up to the first order of $dt$~\cite{zhou2018achieving,layden2019ancilla})
\begin{multline}
\label{eq:effective-1}
\frac{d\rho}{dt} = -i[\omega\mP(H),\rho] + \sum_{i=1}^r \Big( \mP(L_i \rho L_i^\dagger) + \\ \mR(\mP_\perp(L_i \rho L_i^\dagger))  - \frac{1}{2}\{\mP(L_i^\dagger L_i),\rho\}\Big),
\end{multline}
where $P_\perp = 1 - P$, $\mP(\cdot) = P(\cdot)P$, $\mP_\perp(\cdot) = P_\perp(\cdot)P_\perp$ and $\mR$ is a CPTP map describing the AQEC recovery channel. We define the following class of AQEC codes
\begin{equation}
\label{eq:code}
\ket{0_\tl/1_\tl} = \sum_{ij} A_{0/1,ij}\ket{i}_{\mH_S}\ket{j,0/1}_{\mH_A},
\end{equation}
where $A_{0},A_{1}\in \bC^{d\times d}$ and $A_{0/1,ij} = C_{ij} \pm \epsilon D_{ij}$ satisfy $\trace(A_0A_0^\dagger) = \trace(A_1A_1^\dagger) = 1$ and $\trace(C^\dagger D) = 0$.  {Here $C$ describes the part of the code which $\ket{0_\tl}$ and $\ket{1_\tl}$ have in common and $D$ describes the part distinguishing $\ket{0_\tl}$ from $\ket{1_\tl}$ which generates non-zero signal and noise. In the special case where $\epsilon = 0$, the effective signal and noise are zero.} Let $\mH_A = \mH_{A'} \otimes \mH_2 $ where $\dim \mH_{A'} = d$ and $\dim \mH_2 = 2$, the last ancillary qubit in $\mH_2$ makes the signal and noises both diagonal in the code space, i.e. $\braket{0_\tl|H|1_\tl} = \braket{0_\tl|S|1_\tl} = 0$ for all $S \in \mS$. Later on, we will assume $\epsilon$ is a small parameter and consider the perturbation expansion of the effective dynamics around $\epsilon = 0$. We consider the recovery channel restricted to the structure (we will show that this type of recovery channels is sufficient for our purpose)
\begin{multline}
\label{eq:recovery}
\mR(\cdot) = \sum_m \left(\ket{0_\tl}\bra{R_m,0} + \ket{1_\tl}\bra{S_m,1}\right) (\cdot) \\ \left(\ket{R_m,0}\bra{0_\tl} + \ket{S_m,1}\bra{1_\tl}\right),
\end{multline} 
where $\{\ket{R_m}\},\{\ket{S_m}\} \subset \mH_{S}\otimes \mH_{A'}$ are two sets of orthonormal basis and $\mR$ is CPTP. 
A few lines of calculation shows the effective channel (\eqref{eq:effective-1}) under the AQEC code (\eqref{eq:code}) and the recovery channel (\eqref{eq:recovery}) is
\begin{equation}
\label{eq:effective-2}
\frac{d\rho}{dt} \!=\!-\!i\left[\frac{\omega \trace(H Z_\tl)}{2}
 Z_\tl  \!+\! H_{\ts},\rho\right] \!+\! \frac{\gamma(\mR)}{2}\left(Z_\tl \rho Z_\tl \!-\! \rho\right),
\end{equation}
where $Z_\tl = \ket{0_\tl}\bra{0_\tl} - \ket{1_\tl}\bra{1_\tl}$, $H_{\ts}$ is independent of $\omega$, and 
\begin{multline}
\label{eq:gamma-R}
\gamma(\mR) = - \Re\Big[ \sum_{i=1}^r \bra{0_\tl}\Big( \mR(\mP_\perp(L_i \ket{0_\tl}\bra{1_\tl} L_i^\dagger)) +\\ \mP(L_i \ket{0_\tl}\bra{1_\tl} L_i^\dagger)  - \frac{1}{2}\{\mP(L_i^\dagger L_i),\ket{0_\tl}\bra{1_\tl}\}\Big) \ket{1_\tl} \Big]. 
\end{multline}
We can remove the term $H_{\ts}$ in \eqref{eq:effective-2} by applying a reverse Hamiltonian constantly~\cite{sekatski2017quantum}. 
For dephasing channels, the optimal $\frakF$ is reached using a special type of spin-squeezed state as the input~\cite{kitagawa1993squeezed,huelga1997improvement,ulam2001spin,escher2011general,demkowicz2014using}, where we have  
\begin{equation}
\label{eq:sensitivity}
\frakF = \frac{\trace(H Z_\tl)^2}{2\gamma(\mR)}.
\end{equation}
{To simulate the evolution of multipartite spin-squeezed states using the sequential strategy where we have only a single probe, one could first prepare the desired spin-squeezed state in $\bigotimes_{i=1}^N\mH_i$ by entangling the logical qubit in the effective dephasing channel ($\mH_1 = \mH_S \otimes \mH_A$) with a large number of ancillas ($\bigotimes_{i=2}^N\mH_i$) where $\dim \mH_i = \dim \mH_1$ for $2 \leq i \leq N$, and then perform swap operations between $\mH_1$ and $\mH_i$ for $i=2,\ldots,N$ successively every time $t/N$. The optimal $\frakF$ in \eqref{eq:sensitivity} is asymptotically attainable at $N \rightarrow \infty$~\cite{ulam2001spin}. }

For simplicity in furture calculation, we perform a two-step gauge transformation on the Lindblad operators $\{L_i\}_{i=1}^r$ to simplify the dynamics: (1) Let $L_i \leftarrow L_i - \trace(C^\dagger L_i C) \cdot \id$,
such that $L_i$ satisfies $\trace(C^\dagger L_i C) = 0$ for all $L_i$. (2) Perform a unitary transformation $u \in \bC^{r\times r}$ on the Lindblad operators $\vL \leftarrow u \vL,$
such that $\trace(C^\dagger L_i^\dagger L_j C)$ is a diagonal matrix. Note that above transformations only induce another parameter-independent shift $H_\ts$ in the Hamiltonian which could be eliminated by a reverse Hamiltonian. Now we have a new set of Lindblad operators $\{J_i\}_{i=1}^r$, satisfying
\begin{equation}
\label{eq:Jprop}
\trace(C^\dagger J_i C) = 0,
\quad
\trace(C^\dagger J_i^\dagger J_j C) = \lambda_{i}\delta_{ij}, 
\end{equation}
and we replace $\{L_i\}_{i=1}^r$ with $\{J_i\}_{i=1}^r$ in \eqref{eq:gamma-R}. 

First, we maximize $\mathfrak{F}$ over the recovery $\mR$, which is equivalent to minimizing $\gamma(\mR)$ over $\mR$. We claim that the minimum noise rate $\gamma = \min_\mR \gamma (\mR)$ is 
\begin{multline}
\label{eq:gamma}
\gamma = - \Big\| \sum_{i=1}^r \mP_\perp(J_i \ket{0_\tl}\bra{1_\tl} J_i^\dagger) \Big\|_1 - \Re\Big[ \sum_{i=1}^r \bra{0_\tl}\\ \Big( \mP(J_i \ket{0_\tl}\bra{1_\tl} J_i^\dagger)  - \frac{1}{2}\{\mP(J_i^\dagger J_i),\ket{0_\tl}\bra{1_\tl}\}\Big) \ket{1_\tl} \Big],
\end{multline}
where we have used $\max_{U:U^\dagger U = \id}\trace(MU+M^\dagger U^\dagger) = 2\norm{M}_1$ for arbitrary square matrices $M$ and $U$, where $\norm{\cdot}_1$ is the trace norm (see details in \appref{app:gamma}). 

Next, we could like to maximize $\frakF$ over all possible AQEC codes of the form \eqref{eq:code}. It is not clear yet how that could be done mathematically with the presence of trace norm in the denominator. To arrive at an expression of $\gamma$ free of the trace norm, we further sacrifice the generality of our AQEC code and assume $\epsilon \ll 1$. We call it the ``perturbation'' code in the sense that the signal and the noise are both infinitesimally small when $\epsilon \rightarrow 0$. Under the limit $\epsilon \rightarrow 0$, we have $\trace(HZ_\tl) = 2 \epsilon \trace(H\tilde C) + O(\epsilon^2)$, where 
\begin{equation}
\label{eq:tilde-C}
\tilde{C} = C D^\dagger + D C^\dagger,
\end{equation}
and the noise rate is (ignoring all $o(\epsilon^2)$ terms)
\begin{equation}
\label{eq:gamma-E}
\gamma = \epsilon^2 \bigg(\! \sum_i 2 \big|\trace(J_i\tilde{C})\big|^2 \!+\! \sum_{ij:\lambda_i+\lambda_j\neq 0}\!  \frac{|\trace(J_i^\dagger J_j \tilde C)|^2}{(\lambda_i+\lambda_j)} \!\bigg).
\end{equation}
For a detailed derivation of the noise rate, see \appref{app:perturbation} and \cite{zhou2019an}. Finally, we have the following expression of the normalized QFI (up to the lowest order of $\epsilon$) 
\begin{equation}
\label{eq:QFI}
\frakF(C,\tilde C) \!\approx\! \frac{\trace(H \tilde{C})^2}{\sum_i \big|\trace(J_i\tilde{C})\big|^2 \!+\! \sum_{ij:\lambda_i+\lambda_j \neq 0} \!\frac{|\trace(J_i^\dagger J_j \tilde C)|^2}{2(\lambda_i+\lambda_j)}},
\end{equation}
as a function of $\tilde C$ and $C$ (implicitly through the choice of $\{J_i\}_{i=1}^r$). The effective dynamics of the perturbation code has the feature that both the signal and the noises are equally weak and only the ratio between them matters. Therefore the exact value of $\epsilon$ will not influence the normalized QFI $\frakF$ as long as it is sufficiently small. On the other hand, it does influence how fast $F(t)/t$ reaches its optimum $\frakF$, characterized by a coherence time $O(1/\epsilon^2)$.

\paragraph*{Saturating the bound.--} 

Now we maximize the normalized QFI (up to the lowest order of $\epsilon$) over $C$ and $\tilde C$  and show that the optimal $\frakF$ is exactly equal to its upper bound in \eqref{eq:upper}. The domain of $C$ is all complex matrices satisfying $\trace(C^\dagger C) = 1$. We assume the domain of $\tilde C$ is all traceless Hermitian matrices satisfying $\trace(J^\dagger_i J_j \tilde{C})= 0$ for all $i,j \in \inull:=\{i | \lambda_i = 0\}$. When $C$ is full-rank, $\inull$ is empty and for arbitrary traceless $\tilde C$, we could always take $D^\dagger = \frac{1}{2} C^{-1} \tilde C$ such that \eqref{eq:tilde-C} is satisfied. When $C$ is singular, we could replace it with an approximate full-rank version (e.g. $C \leftarrow C + \delta \id$). In this case, $\frakF$ will only be decreased by an infinitesimal small amount when $\epsilon = o(\delta^2)$ because the numerator in \eqref{eq:QFI} is only slightly perturbed after the replacement. 

Consider the following optimization problem over $h,\vh,\frakh$ and $C$,
\begin{equation}
\label{eq:reach}
\begin{split}
&\max_{C} \min_{h,\vh,\frakh}~~  4 \trace(C^\dagger \alpha C) 
,\\
&\;\text{subject~to~}~ \beta = 0,\quad\trace(C^\dagger C)= 1,
\end{split}
\end{equation}
Fixing $C$, we introduce a Hermitian matrix $\tilde C$ as the Lagrange multiplier associated with the constraint $\beta = 0$~\cite{boyd2004convex}. Strong duality implies \eqref{eq:reach} has the same solution as the following dual program (see \appref{app:dual})
\begin{equation}
\label{eq:dual}
\begin{split}
\max_{C,\tilde C}~ \frakF(C,\tilde C),&\text{~~subject~to~}\trace(C^\dagger C)= 1,\,\trace(\tilde{C}) = 0,\\
&~~\text{~and~}\trace(J^\dagger_i J_j \tilde{C}) = 0,\forall i,j \in \inull. 
\end{split}
\end{equation}
whose optimal value could be achieved using the perturbation code up to an infinitesimal small error according to the discussion above. On the other hand, thanks to Sion's minimax theorem~\cite{komiya1988elementary,do2001introduction}, we can exchange the order of the maximization and minimization in \eqref{eq:reach} because we could always confine $(\vh,\frakh)$ in a convex and compact set (see \appref{app:compact}) such that the solution of \eqref{eq:reach} is not altered and the objective function $4\trace(C^\dagger \alpha C)$ is concave (linear) with respect to $CC^\dagger$ and convex (quadratic) with respect to $(\vh,\frakh)$. Therefore, the optimal value of \eqref{eq:dual} is also equal to $4\min_{h,\vh,\frakh|\beta = 0}\norm{\alpha}$, the upper bound of the normalized QFI.

\paragraph*{Numerical algorithm.--}

It is known that the upper bound in \eqref{eq:upper} could be calculated via a semidefinite program (SDP)~\cite{demkowicz2017adaptive,czajkowski2019many},
\begin{equation}
\label{eq:SDP}
F(t) \leq 4t \min_{h,\vh,\frakh|\beta = 0} x, \text{~subject~to~} \scrA \succeq 0,\;\beta = 0, 
\end{equation}
where $\scrA = \begin{pmatrix} \sqrt{x} \id & \vh^\dagger \id + \vL^\dagger \frakh \\ \vh \id + \frakh \vL & \sqrt{x} \id^{\otimes r}\\\end{pmatrix}$ and ``$\succeq 0$'' means positive semidefinite.
However, the minimax theorem does not guaranteed an efficient algorithm to solve \eqref{eq:dual} after exchanging the order of the maximization and minimization in \eqref{eq:reach}. Now we provide an efficient numerical algorithm obtaining an optimal $(C^\diamond,\tilde C^\diamond)$ in three steps. The validity of this algorithm is proven in \appref{app:algorithm}. The algorithm runs as follows: (a)~Solving $\min_{h,\vh,\frakh|\beta = 0}\norm{\alpha}$ using the SDP gives us an optimal $\alpha^\diamond$ (and corresponding $h^\diamond,\vh^\diamond,\frakh^\diamond$) satisfying $\norm{\alpha^\diamond} = \min_{h,\vh,\frakh|\beta = 0}\norm{\alpha}$. (b)~Suppose $\Pi^\diamond$ is the projection onto the subspace spanned by all eigenstates corresponding to the largest eigenvalue of $\alpha^\diamond$, we find an optimal $C^\diamond C^{\diamond\dagger}$ satisfying $\Pi^\diamond C^\diamond C^{\diamond\dagger} \Pi^\diamond= C^\diamond C^{\diamond\dagger}$ and  
\begin{equation} 
\Re[\trace(C^\diamond C^{\diamond\dagger}(\dvh \id + \dfrakh \vL)^\dagger(\vh^\diamond\id + \frakh^\diamond \vL))] = 0,
\end{equation}
for all $(\dvh,\dfrakh)$ such that $\Delta h\id + \dvh^\dagger \vL + \vL^\dagger \dvh + \vL^\dagger \dfrakh \vL = 0$ for some $\Delta h$. Note that this step is simply solving a system of linear equations. (c) Find $\{J_i\}_{i=1}^r$ via the gauge transformation. Let $\mS_0 = {\rm span}\{I,J_i^\dagger J_j,\forall i,j\in\inull\}$. Decompose $M = J_i$ or $J_{ij}$ ($:=J_i^\dagger J_j$) into $M = M^{h} + i M^{ah} + M_0^{h} + i M_0^{ah}$ where $M^{h,ah}$, $M_0^{h,ah}$ are Hermitian, $M_0^{h,ah} \in \mS_0$ and $M^{h,ah} \perp \mS_0$ (in terms of the Hilbert-Schmidt norm). Using the vectorization of matrices $\dket{\cdot} = \sum_{jk} \bra{j}(\cdot)\ket{k} \ket{j}\ket{k}$, let 
\begin{multline}
B = \sum_i \dket{J_i^h}\dbra{J_i^h} + \dket{J_i^{ah}}\dbra{J_i^{ah}}  + \\ \sum_{ij:\lambda_i+\lambda_j \neq 0} \frac{\dket{J_{ij}^h}\dbra{J_{ij}^h} + \dket{J_{ij}^{ah}}\dbra{J_{ij}^{ah}}}{2(\lambda_i+\lambda_j)}.
\end{multline}
According to the Cauchy-Schwarz inequality, 
\begin{equation}
\max_{\tilde C} \frakF(C^\diamond,\tilde C) = \max_{\tilde C} \frac{|\dbraket{H|\tilde{C}}|^2}{\dbraket{\tilde{C}|B|\tilde{C}}}= \dbraket{H^{h}|B^{-1}|H^{h}},
\end{equation}
and the optimal $\dket{\tilde C^\diamond} = B^{-1} \dket{H^h}$. Here $^{-1}$ means the Moore-Penrose pseudoinverse.

\paragraph*{Highly-biased noise.--}

We consider a special case where noises are separated into two groups -- strong ones and weak ones~\cite{zhou2018achieving,layden2018spatial,layden2019ancilla}. To be specific, we consider the following quantum master equation
\begin{multline}
\label{eq:biased}
\frac{d\rho}{dt} = -i[\omega H,\rho] + \sum_{i\in\iw} \eta \Big(L_i \rho L_i^\dagger - \frac{1}{2}\{L_i^\dagger L_i,\rho\}\Big) \\+ \sum_{i\in\is} \Big(L_i \rho L_i^\dagger - \frac{1}{2}\{L_i^\dagger L_i,\rho\}\Big),
\end{multline}
where the indices of Lindblad operators $\{L_i\}_{i=1}^r$ are separated into $\iw$ and $\is$, representing weak and strong noises respectively. $\eta \ll 1$ is a small parameter characterizing the relative strength of the weak noises. Moreover, we assume that $H \notin {\rm span}\{\id,L_i,L_i^\dagger,L_i^\dagger L_j,\,i,j\in\is\}$ so that it is possible to fully correct all strong noises and also preserve a non-trivial signal in the code space. Taking $\eta \rightarrow 0$, it is easy to show that (see \appref{app:biased}), the optimal $\frakF$ in this case is equal to 
\begin{equation}
\label{eq:biased-QFI}
\frakF^\diamond = \frac{1}{\eta}\norm{\balpha}_{h,\vh,\frakh|\beta = 0} + O(1),
\end{equation}
where $\balpha = (\vh\id + \frakh \vL)^\dagger \Pi_\iw (\vh\id + \frakh \vL)$ and $\Pi_\iw$ is a diagonal matrix whose $i$-th diagonal element is one when $i\in\iw$ and zero when $i\in\is$. 
This reduces the running time of the SDP in \eqref{eq:SDP} by reducing $\scrA$ from a $d(r+1) \times d(r+1)$ matrix to a $d(\abs{\iw}+1) \times d(\abs{\iw}+1)$ matrix. Using the optimal AQEC strategy, $\frakF$ is boosted by a factor of $O(1/\eta)$, compared to the case where no QEC is performed. To find the optimal AQEC code, we can solve the dual program of a modified version of \eqref{eq:reach} where $\alpha$ is replaced with $\balpha$: 
\begin{align}
\max_{C,\tilde C}~ \overline{\frakF}(C,\tilde C),&\text{~~subject~to~}\trace(C^\dagger C)= 1,\,\trace(\tilde{C}) = 0,\\
&~~\forall_{i,j \in \is},\;\trace(L^\dagger_i L_j \tilde{C}) = \trace(L_i\tilde{C}) = 0\label{eq:KL},
\end{align}
and some other linear contraints on $\tilde{C}$ when $\trace(C \vL^\dagger \vL C) - \trace(C \vL^\dagger C)\trace(C \vL C)$ is singular. Here $\overline{\frakF}$ is the dominant part of $\frakF$ such that 
$\frakF = \overline{\frakF}/\eta + O(1)$.
Detailed calculations including the expression of $\overline{\frakF}$ and the dual program are provided in \appref{app:biased}. The constraint \eqref{eq:KL} on $\tilde C$ is equivalent to the Knill-Laflamme condition for Lindblad operators $\{L_i\}_{i\in\is}$~\cite{knill1997theory,beny2011perturbative}. It implies strong noises are fully corrected by the optimal AQEC code and explains why the estimation precision depends only on the strength of weak noises in~\eqref{eq:biased-QFI}.

\paragraph*{Conclusions and outlook.--} In this Letter, we proposed an AQEC strategy such that the optimal SQL in Hamiltonian parameter estimation could be achieved asymptotically. An interesting open question is whether the perturbation code we introduced here could be turned into non-pertubative ones. We provide an example in \appref{app:dephasing}, where by slightly modifying the ancilla-free QEC code (non-perturbation) proposed in Ref.~\cite{layden2019ancilla}, we show that the optimal $\frakF$ could be achieved in the correlated dephasing noise model. However, it is unclear how to generalize the result to generic noise models. Another two interesting open questions are (1) how to characterize the power of QEC in improving quantum metroloy for parameters encoded in generic quantum channels~\cite{demkowicz2012elusive,demkowicz2014using}, for example when the rate of quantum controls is constant, rather than infinitely fast; (2) how to optimize the QEC strategy when considering a constant probing time, rather than an infinitely long probing time.

\paragraph*{Acknowledgements.--} We thank Kyungjoo Noh, Rafa{\l} Demkowicz-Dobrza\'{n}ski, Zhou Fan, Jing Yang, Yuxiang Yang for helpful discussions. We acknowledge support from the ARL-CDQI (W911NF15-2-0067, W911NF-18-2-0237), ARO (W911NF-18-1-0020, W911NF-18-1-0212), ARO MURI (W911NF-16-
1-0349), AFOSR MURI (FA9550-15-1-0015), DOE (DE-SC0019406), NSF (EFMA-1640959), and the Packard Foundation (2013-39273).

\newpage

\bibliographystyle{aps}

\onecolumngrid
\newpage
\appendix

\section{\label{app:gamma}\texorpdfstring{Minimizing the noise rate $\gamma(\mR)$ over recovery channels $\mR$}{Minimizing the noise rate over recovery channels}}

In this appendix we prove \eqref{eq:gamma} in the main text. According to \eqref{eq:gamma-R}, 
\begin{equation}
\gamma(\mR) = - \Re\Big[ \sum_{i=1}^r \bra{0_\tl}\Big( \mR(\mP_\perp(J_i \ket{0_\tl}\bra{1_\tl} J_i^\dagger)) + \mP(J_i \ket{0_\tl}\bra{1_\tl} J_i^\dagger)  - \frac{1}{2}\{\mP(J_i^\dagger J_i),\ket{0_\tl}\bra{1_\tl}\}\Big) \ket{1_\tl} \Big]. 
\end{equation}
In order to calculate $\gamma = \min_\mR \gamma(\mR)$, we only need to calculate the first term minimized over $\mR$:
\begin{equation}
\begin{split}
&\quad -\max_\mR \Re\bigg[ \sum_i \bra{0_\tl}\mR(\mP_\perp(J_i \ket{0_\tl}\bra{1_\tl} J_i^\dagger))\ket{1_\tl}\bigg] \\
&= - \max_{\ket{R_m},\ket{S_m}} \Re\bigg[  \sum_{i,m}\bra{R_m,0}P_\perp J_i\ket{0_\tl}\bra{1_\tl}J_i^\dagger P_\perp \ket{S_m,1} \bigg]\\
&= - \frac{1}{2}\max_{\ket{R_m},\ket{S_m}} \trace\Big(\sum_m \ket{R_m}\bra{S_m} \cdot \sum_i \bra{0}P_\perp J_i\ket{0_\tl}\bra{1_\tl}J_i^\dagger P_\perp\ket{1} + h.c. \Big)\\
&= - \Big\| \sum_i \bra{0}P_\perp J_i\ket{0_\tl}\bra{1_\tl}J_i^\dagger P_\perp\ket{1}\Big\|_1 = - \Big\| \sum_i  P_\perp J_i\ket{0_\tl}\bra{1_\tl}J_i^\dagger P_\perp \Big\|_1,\\
\end{split}
\end{equation}
where $h.c.$ means Hermitian conjugate and we have used $\max_{U:U^\dagger U = \id}\trace(MU+M^\dagger U^\dagger) = 2\norm{M}_1$ for arbitrary square matrices $M$ and $U$, which could be proven easily using the singular value decomposition of $M$.

\section{\label{app:perturbation}\texorpdfstring{Perturbation expansion of the noise rate $\gamma$}{Perturbative expansion of the noise rate}}

In this appendix we expand the minimum noise rate $\gamma$ around $\epsilon = 0$ using the perturbation code. For simplicity, the \emph{equal sign} ``$=$'' in this appendix means approximate equality up to the second order of $\epsilon$ (ignoring all $o(\epsilon^2)$ terms). We also state a useful lemma here:
\begin{lemma}[\cite{mirsky1960symmetric}]
$\norm{X+ \epsilon Y}_1 = \norm{X}_1 + O(\epsilon)$ for arbitrary $X$ and $Y$.
\end{lemma}


To calculate \eqref{eq:gamma}, we first consider the terms independent of $\mR$, 
\begin{equation}
\label{eq:first-term}
\begin{split}
&\quad - \Re\left[ \sum_{i=1}^r \bra{0_\tl}\left( \mP(J_i \ket{0_\tl}\bra{1_\tl} J_i^\dagger) - \frac{1}{2}\{\mP(J_i^\dagger J_i),\ket{0_\tl}\bra{1_\tl}\}\right) \ket{1_\tl} \right]\\
&= - \sum_{i} \Re\big[\trace(A_0A_0^\dagger J_i)\trace(A_1A_1^\dagger J_i^\dagger)]-\frac{1}{2}\left(\trace(A_0A_0^\dagger J_i^\dagger J_i) + \trace(A_1A_1^\dagger J_i^\dagger J_i)\right) \\
&= \sum_{i}  \lambda_i + \epsilon^2 \big|\trace(\tilde C J_i)\big|^2 + \epsilon^2 \trace(DD^\dagger J_i^\dagger J_i). 
\end{split}
\end{equation}
The remaining term is equal to (thanks to Lemma 1) minus
\begin{equation}
\label{eq:nuclear-norm}
\Big\| \sum_i  P_\perp J_i\ket{0_\tl}\bra{1_\tl}J_i^\dagger P_\perp \Big\|_1 
= 
\left\| 
\begin{pmatrix}
\sqrt{\tLambda^{-1}}(\Lambda+\epsilon X_1+\epsilon^2 X_1')\\
\epsilon X_2+\epsilon^2 X_2'
\end{pmatrix}
\begin{pmatrix}
(\Lambda-\epsilon X_1+\epsilon^2 X_1')^\dagger\sqrt{\tLambda^{-1}}& -\epsilon X_2^\dagger+\epsilon^2 X_2'^\dagger
\end{pmatrix}
 \right\|_1,
\end{equation}
where $\Lambda \in \bR^{r\times r}$ is a diagonal matrix whose $k$-th diagonal element is $\lambda_k$ and $\tLambda \in \bR^{r\times r}$ is a diagonal matrix whose $k$-th diagonal element is $\lambda_k$ if $\lambda_k > 0$ and $1$ if $\lambda_k = 0$. Assume $\{\lambda_k\}_{k=1}^r$ is arranged in a non-ascending order and $r_0$ is the largest integer such that $\lambda_{r_0}$ is positive. $X_1,X_1' \in \bC^{r\times r}$ satisfy
\begin{equation}
\begin{split}
(\Lambda+\epsilon X_1 + \epsilon^2 X_1')_{ji} 
&= \sqrt{\lambda_j}\bra{\tilde{J}_{j,0}}P_\perp J_i \ket{0_\tl} = \trace(C^\dagger J_j^\dagger J_i A_0) - \trace(C^\dagger  J_j^\dagger  A_0)\trace(A_0^\dagger J_i A_0) \\
&= \lambda_{i}\delta_{ij} + \epsilon \trace(C^\dagger J_j^\dagger J_i D) - \epsilon^2 \trace(C^\dagger J_j^\dagger D)\trace(\tilde C J_i),
\end{split}
\end{equation}
for $1 \leq j \leq r_0$ and
\begin{equation}
\begin{split}
(\Lambda+\epsilon X_1 + \epsilon^2 X_1')_{ji} 
&= \bra{\tilde{J}_{j,0}}P_\perp J_i \ket{0_\tl} = \trace(\tJ_{j}^\dagger J_i A_0) - \trace(\tJ_{j}^\dagger  A_0)\trace(A_0^\dagger J_i A_0) \\
&=  \epsilon \trace(\tJ_{j}^\dagger J_i D) - \epsilon^2 \trace(\tJ_{j}^\dagger  D) \trace(\tilde C J_i),
\end{split}
\end{equation}
for $r_0+1 \leq j \leq r$. 
$X_2,X_2' \in \bC^{(d^2-r)\times r}$ satisfy 
\begin{equation}
\begin{split}
(\Lambda+\epsilon X_2 + \epsilon^2 X_2')_{ji} 
&= \bra{\tilde{J}_{j+r,0}}P_\perp J_i \ket{0_\tl} = \trace(\tJ_{j+r}^\dagger J_i A_0) - \trace(\tJ_{j+r}^\dagger  A_0)\trace(A_0^\dagger J_i A_0) \\
&=  \epsilon \trace(\tJ_{j+r}^\dagger J_i D) - \epsilon^2 \trace(\tJ_{j+r}^\dagger  D) \trace(\tilde C J_i),
\end{split}
\end{equation}
for $r +1 \leq j \leq d^2 - 1$ and 
\begin{equation}
\begin{split}
(\Lambda+\epsilon X_2 + \epsilon^2 X_2')_{d^2 i} 
&= \bra{\tilde{J}_{d^2,0}}P_\perp J_i \ket{0_\tl} = \trace(\tJ_{d^2}^\dagger J_i A_0) - \trace(\tJ_{d^2}^\dagger  A_0)\trace(A_0^\dagger J_i A_0) \\
&=  \epsilon \trace(C^\dagger J_i D) - \epsilon \trace(\tilde C J_i). 
\end{split}
\end{equation}
Here  
\begin{equation}
\ket{\tilde{J}_{j,0/1}} = 
\begin{cases}
\frac{1}{\sqrt{\lambda_j}}\sum_{ik} C_{ik} J_j \ket{i}\ket{k,0/1}, & j \leq r_0,\\
\sum_{ik} (\tJ_j)_{ik} \ket{i}\ket{k,0/1},  & r_0 < j \leq d^2,\\
\sum_{ik} \frac{C_{ik}}{\sqrt{\trace(C^\dagger C)}} \ket{i}\ket{k,0/1},  &  j = d^2,\\
\end{cases}
\end{equation}
are two sets of orthonormal basis of $\mH_S \otimes \mH_A$. 

To calculate the first and second order expansion of \eqref{eq:nuclear-norm}, we consider the singular value decompositions
\begin{equation}
\begin{split}
\begin{pmatrix}
\sqrt{\tLambda^{-1}}(\Lambda+\epsilon X_1+\epsilon^2 X_1')\\
\epsilon X_2+\epsilon^2 X_2'
\end{pmatrix} &= U(\epsilon) \begin{pmatrix}\Sigma(\epsilon)\\0\end{pmatrix} V(\epsilon)^\dagger,\\
\begin{pmatrix}
(\Lambda-\epsilon X_1+\epsilon^2 X_1')^\dagger\sqrt{\tLambda^{-1}}& -\epsilon X_2^\dagger+\epsilon^2 X_2'^\dagger
\end{pmatrix} &= V(-\epsilon) \begin{pmatrix}\Sigma(-\epsilon)&0\end{pmatrix} U(-\epsilon)^\dagger,
\end{split}
\end{equation}
Then 
\begin{equation}
\begin{split}
\text{\eqref{eq:nuclear-norm}} &= \norm{ U(\epsilon) \begin{pmatrix}
\Sigma(\epsilon)V(\epsilon)^\dagger V(-\epsilon)\Sigma(-\epsilon) & 0 \\
0 & 0
\end{pmatrix} U(-\epsilon)^\dagger}_1 = \norm{\Sigma(\epsilon)V(\epsilon)^\dagger V(-\epsilon)\Sigma(-\epsilon)}_1\\
&= \trace\left(\big(V(\epsilon)\Sigma(\epsilon)V(\epsilon)^\dagger V(-\epsilon)\Sigma(-\epsilon)\Sigma(-\epsilon)V(-\epsilon)^\dagger V(\epsilon)\Sigma(\epsilon)V(\epsilon)^\dagger\big)^{1/2}\right)\\
&= \trace\left( \sqrt{\sqrt{Y(\epsilon)} Y(-\epsilon) \sqrt{Y(\epsilon)}}\right),
\end{split}
\end{equation}
where 
\begin{equation}
\begin{split}
Y(\epsilon) = V(\epsilon)\Sigma(\epsilon)^2V(\epsilon)^\dagger 
&= \Lambda + \epsilon(X_1^\dagger \Pi_{\Lambda}  + \Pi_{\Lambda} X_1) + \epsilon^2 (X_1^\dagger \tLambda^{-1} X_1 + X_1' \Pi_{\Lambda}  + \Pi_{\Lambda} X_1'^\dagger + X_2^\dagger X_2),\\
&=: \Lambda + \epsilon W + \epsilon^2 W'
\end{split}
\end{equation}
and $\Pi_{\Lambda}$ is the projector onto the support of $\Lambda$. 

Using Theorem 2 in Ref.~\cite{zhou2019an}, we have
\begin{multline}
\trace\left( \sqrt{\sqrt{Y(\epsilon)} Y(-\epsilon) \sqrt{Y(\epsilon)}}\right) = \\
\trace(\Lambda) + \epsilon^2 \trace(X_1^\dagger \tLambda^{-1} X_1 + X_1' \Pi_{\Lambda}  + \Pi_{\Lambda} X_1'^\dagger + X_2^\dagger X_2) - \epsilon^2 \sum_{i,j:\lambda_i+\lambda_j\neq 0}^r \frac{|X_{1,ij}+X_{1,ji}^*|^2}{\lambda_i+\lambda_j}.
\end{multline}
Note that 
\begin{gather}
\trace(X_1^\dagger \tLambda^{-1} X_1  + X_2^\dagger X_2) = \sum_{i=1}^r \trace(DD^\dagger J_i^\dagger J_i) + \sum_{i=1}^{r}\abs{\trace(\tilde C J_i)}^2 - \sum_{i=1}^{r} \trace(D^\dagger J_i^\dagger C)\trace(\tilde C J_i) + \trace(C^\dagger J_i D)\trace(\tilde C J_i^\dagger),\\
\trace(X_1' \Pi_{\Lambda}  + \Pi_{\Lambda} X_1'^\dagger) = -\sum_{i=1}^{r}\left( \trace(C^\dagger J_i^\dagger D)\trace(\tilde C J_i) + \trace(D^\dagger J_i C)\trace(\tilde C J_i^\dagger) \right),
\end{gather}
therefore
\begin{equation}
\begin{split}
\gamma &= \text{\eqref{eq:first-term}} - \text{\eqref{eq:nuclear-norm}}\\
&= 2{\epsilon^2} \sum_i \big|\trace(J_i\tilde{C})\big|^2 + {\epsilon^2} \sum_{ij:\lambda_i+\lambda_j\neq 0} \frac{|\trace(J_i^\dagger J_j \tilde C)|^2}{(\lambda_i+\lambda_j)},
\end{split}
\end{equation}
where $\tilde{C} = C D^\dagger + D C^\dagger$.
According to \eqref{eq:sensitivity}, we have
\begin{equation}
\frakF = \frac{\trace(H \tilde{C})^2}{\sum_i \big|\trace(J_i\tilde{C})\big|^2 + \sum_{ij:\lambda_i+\lambda_j\neq 0} \frac{|\trace(J_i^\dagger J_j \tilde C)|^2}{2(\lambda_i+\lambda_j)}} + O(\epsilon).
\end{equation}

\section{\label{app:dual}\texorpdfstring{Lagrange dual program of \eqref{eq:reach}}{Lagrange dual program of Eq.~(17)}}

Here we show the Lagrange dual program of \eqref{eq:reach} is \eqref{eq:dual}. 
From the definition of $\alpha$ (\eqref{eq:alpha}) and $\beta$ (\eqref{eq:beta}), we see that the upper bound in \eqref{eq:upper} is invariant under the transformation $\vL \rightarrow \vJ$, that is, after the transformation $\vL \rightarrow \vJ$ there is always another set of $(h,\vh,\frakh)$ such that $\beta = 0$ and $\alpha$ is the same. Therefore we let 
\begin{align}
\alpha &= (\vh\id + \frakh \vJ)^\dagger (\vh\id + \frakh \vJ), 
\\
\beta &= H + h\id + \vh^\dagger \vJ + \vJ^\dagger \vh + \vJ^\dagger \frakh \vJ,
\end{align}
where $\vJ = (J_1,J_2,\ldots,J_r)^T$. To proceed, we simplify the notations by letting
\begin{equation}
\vj_i = \frac{\trace(J_i \tilde C)}{\trace(H\tilde C)},
\quad
\frakj_{ij} = \frac{\trace(J_i^\dagger J_j \tilde C)}{\trace(H\tilde C)}.
\end{equation}
Note that the $r$-dimensional vector $\vj$ is to be distinguished from the index $j$, then we have 
\begin{equation}
\frakF(C,\tilde C) = \bigg(\vj^\dagger \vj + \sum_{ij:\lambda_i+\lambda_j \neq 0} \frac{\abs{\frakj_{ij}}^2}{2(\lambda_i + \lambda_j)}\bigg)^{-1}, 
\end{equation}
and $4\trace(C^\dagger \alpha C) = 4(\vh^\dagger \vh + \trace(\Lambda \frakh^2))$. 

Fixing $C$, we introduce a Hermitian matrix $\tilde C$ as a Lagrange multiplier of $\beta = 0$~\cite{boyd2004convex}, the Lagrange function is 
\begin{equation}
L(\tilde C,h,\vh,\frakh) = 4(\vh^\dagger \vh + \trace(\Lambda \frakh^2)) + \trace(\tilde C (H + h \id + \vJ^\dagger \vh + \vh^\dagger \vJ + \vJ^\dagger \frakh \vJ)). 
\end{equation}
Then the dual program of \eqref{eq:reach} is 
\begin{equation}
\begin{split}
&\quad~ \max_{\tilde C} \min_{h,\vh,\frakh} L(\tilde C,h,\vh,\frakh) \\
&= \max_{\tilde C} \min_{h,\vh,\frakh} 4(\vh^\dagger \vh + \trace(\Lambda \frakh^2)) + \trace(\tilde C (H + h \id + \vJ^\dagger \vh + \vh^\dagger \vJ + \vJ^\dagger \frakh \vJ))\\
&= \max_{\substack{\tilde C:\trace(\tilde C) = 0,\\\trace(\tilde C H) \neq 0}} \min_{\vh,\frakh} 4(\vh^\dagger \vh + \trace(\Lambda \frakh^2)) + \trace(\tilde C H) (1 + \vh^\dagger \vj + \vj^\dagger \vh + \trace(\frakh^T \frakj))  \\
&=  \max_{\substack{\tilde C:\trace(\tilde C) = 0,\\\forall_{i,j\in\inull}\trace(\tilde C J_i^\dagger J_j) = 0,\\\trace(\tilde C H) \neq 0\\}} - \frac{1}{4} \trace(\tilde C H)^2 \vj^\dagger \vj - \frac{1}{8} \trace(\tilde C H)^2 \sum_{ij:\lambda_i+\lambda_j \neq 0} \frac{\abs{\frakj_{ij}}^2}{\lambda_i + \lambda_j} + \trace(\tilde C H) \\
&=  \max_{\substack{\tilde C:\trace(\tilde C) = 0,\\\forall_{i,j\in\inull}\trace(\tilde C J_i^\dagger J_j) = 0,\\}}  \bigg(\vj^\dagger \vj + \sum_{ij:\lambda_i+\lambda_j \neq 0} \frac{\abs{\frakj_{ij}}^2}{2(\lambda_i + \lambda_j)}\bigg)^{-1} = \max_{\substack{\tilde C:\trace(\tilde C) = 0,\\\forall_{i,j\in\inull}\trace(\tilde C J_i^\dagger J_j) = 0}} \frakF(C,\tilde C),  
\end{split}
\end{equation}
as in \eqref{eq:dual}.

\section{\label{app:compact}
\texorpdfstring{Confining $(\vh,\frakh)$ in a compact set}{Confining (h,h) in a compact set}}

The minimax theorem~\cite{do2001introduction} states that for convex compact sets $P \subset \bR^{m}$ and $Q \subset \bR^{n}$ and $f:P\times Q \rightarrow \bR$ such that $f(x,y)$ is a continuous convex (concave) function in $x$ ($y$) for every fixed $y$ ($x$), then 
\begin{equation}
\max_{y \in Q}\min_{x \in P} f(x,y) = \min_{x \in P} \max_{y \in Q} f(x,y).
\end{equation}
In \eqref{eq:reach}, the operator $CC^\dagger$ satisfying $\trace(CC^\dagger) = 1$ is contained in a convex compact set, but the domain of $(h,\vh,\frakh)$ is not compact. Here we show that we could always confine $(\vh,\frakh)$ in a convex and compact set such that the solution of \eqref{eq:reach} is not altered. First we note that $\min_{h,\vh,\frakh|\beta = 0}\norm{\alpha} = a < \infty$ when $H \in \mS$. Note that 
\begin{equation}
\norm{\alpha} = 
\norm{
\begin{pmatrix}
\vh_1 \id + \sum_{i=1}^r \frakh_{1i} L_i \\
\vh_2 \id + \sum_{i=1}^r \frakh_{2i} L_i \\
\vdots \\
\vh_r \id + \sum_{i=1}^r \frakh_{ri} L_i  \\ 
\end{pmatrix}}^2.
\end{equation}
It is clear that there exists some $b > 0$ such that for all $\norm{(\vh,\frakh)}_2 > b$ ($\norm{\cdot}_2$ is the Euclidean norm), we have $\norm{\alpha} > a$. 
Therefore it is easy to find some $b' > 0$ such that 
\begin{equation}
\begin{split}
&\min_{\vh,\frakh} \max_{C} ~~  4 \trace(C^\dagger \alpha C) 
,\\
&\;\text{subject~to~}~ \exists h,\beta = 0,\quad\trace(C^\dagger C)= 1,\quad \norm{(\vh,\frakh)}_2 \leq b',
\end{split}
\end{equation}
and
\begin{equation}
\begin{split}
&\max_{C} \min_{\vh,\frakh}  ~~  4 \trace(C^\dagger \alpha C) 
,\\
&\;\text{subject~to~}~ \exists h,\beta = 0,\quad\trace(C^\dagger C)= 1,\quad \norm{(\vh,\frakh)}_2 \leq b',
\end{split}
\end{equation}
has the same optimal value equal to $4a$, and there exists a saddle point $(\vh^*,\frakh^*,C^*)$ such that 
\begin{equation}
\label{eq:saddle}
\trace(C^{\dagger} \alpha^* C) \leq \trace(C^{*\dagger} \alpha^* C^*) \leq \trace(C^{*\dagger} \alpha C^*)
\end{equation} 
for all $(\vh,\frakh,C)$ satisfying $\exists h,\beta = 0, \trace(C^\dagger C) = 1,$ and $\norm{(\vh,\frakh)}_2 \leq b'$, where $\alpha^* = (\vh^*\id + \frakh^* \vL)^\dagger (\vh^*\id + \frakh^* \vL)$. Moreover, based on the above discussion, $(\vh^*,\frakh^*)$ is not on the boundary, i.e. $\norm{(\vh^*,\frakh^*)}_2 < b'$. The second inequality in \eqref{eq:saddle} is then equivalent to 
\begin{equation}
\Re[\trace(C^{*\dagger}(\dvh \id + \dfrakh \vL)^\dagger(\vh^*\id + \frakh^* \vL)C^*)] = 0,
\end{equation}
for all $(\dvh,\dfrakh)$ satisfying
\begin{equation}
\Delta h\id + \dvh^\dagger \vL + \vL^\dagger \dvh + \vL^\dagger \dfrakh \vL = 0
\end{equation}
for some $\Delta h$. 
Therefore, $(\vh^*,\frakh^*,C^*)$ is also a saddle point of \eqref{eq:reach}:
\begin{equation}
\begin{split}
&\max_{C} \min_{\vh,\frakh}  ~~  4 \trace(C^\dagger \alpha C) 
,\\
&\;\text{subject~to~}~ \exists h,\beta = 0,\quad\trace(C^\dagger C)= 1,
\end{split}
\end{equation}
proving that the optimal value of \eqref{eq:reach} must also be equal to $4a$.

\section{\label{app:algorithm}The validity of the numerical algorithm}

Here we prove the validity of the three-step algorithm introduced in the main text.
Let $(\vh^*,\frakh^*,C^*)$ be the saddle point of \eqref{eq:reach}. The first inequality in \eqref{eq:saddle} implies
\begin{equation}
\trace(C^{*\dagger} \alpha^* C^*) = \norm{\alpha^*} = \min_{h,\vh,\frakh|\beta = 0}\norm{\alpha},
\end{equation}
which means that $\Pi^* C^* = C^*$ where $\Pi^*$ is the projection onto the subspace spanned by all eigenstates corresponding to the largest eigenvalue of $\alpha^*$. 

Now assume we have a solution $(\vh^\diamond,\frakh^\diamond)$ of \eqref{eq:upper} such that $\alpha^\diamond = (\vh^\diamond\id + \frakh^\diamond \vL)^\dagger (\vh^\diamond\id + \frakh^\diamond \vL)$ satisfies 
\begin{equation}
\norm{\alpha^\diamond} = \min_{h,\vh,\frakh|\beta = 0}\norm{\alpha}. 
\end{equation}
We prove that $(\vh^\diamond,\frakh^\diamond,C^*)$ is also a saddle point. Choose $p \in (0,1)$ and let 
\begin{equation}
(\vh,\frakh) = (p\vh^\diamond + (1-p) \vh^*,p\frakh^\diamond + (1-p) \frakh^*).
\end{equation} 
Then 
\begin{equation}
\label{eq:cauchy}
\begin{split}
\trace(C^{*\dagger} \alpha C^*) 
&= p^2 \trace(C^{*\dagger} \alpha^\diamond C^*) + (1-p)^2 \trace(C^{*\dagger} \alpha^* C^*) + 2 p(1-p) \Re[\trace(C^{*\dagger} (\vh^\diamond\id + \frakh^\diamond \vL)^\dagger (\vh^*\id + \frakh^* \vL)  C^*)]\\
&\leq p^2 \trace(C^{*\dagger} \alpha^\diamond C^*) + (1-p)^2 \trace(C^{*\dagger} \alpha^* C^*) + 2 p(1-p)\sqrt{\trace(C^{*\dagger} \alpha^\diamond C^*)\trace(C^{*\dagger} \alpha^* C^*)}\leq \norm{\alpha^*}.
\end{split}
\end{equation}
On the other hand, we know $\trace(C^{*\dagger} \alpha C^*) \geq \norm{\alpha^*}$. Therefore the equality in \eqref{eq:cauchy} must hold, which means 
\begin{equation}
\trace(C^{*\dagger} \alpha^\diamond C^*) = \norm{\alpha^\diamond}, \quad (\vh^*\id + \frakh^* \vL) C^* = (\vh^\diamond\id + \frakh^\diamond \vL) C^*. 
\end{equation}
As a result, we have $\trace(C^{\dagger} \alpha^\diamond C) \leq \trace(C^{*\dagger} \alpha^\diamond C^*)$ for arbitrary $C$ satisfying $\trace(C^\dagger C) = 1$. Moreover, 
\begin{equation}
\Re[\trace(C^{*\dagger}(\dvh \id + \dfrakh \vL)^\dagger(\vh^\diamond\id + \frakh^\diamond \vL)C^*)] = \Re[\trace(C^{*\dagger}(\dvh \id + \dfrakh \vL)^\dagger(\vh^* \id + \frakh^* \vL)C^*)] = 0,
\end{equation}
and $\trace(C^{*\dagger} \alpha^\diamond C^*) \leq \trace(C^{*\dagger} \alpha C^*)$, proving $(\vh^\diamond,\frakh^\diamond,C^*)$ is also a saddle point. Hence, step (b) in our algorithm will at least have one solution $C^*$, and the solution of step (b) $(\vh^\diamond,\frakh^\diamond,C^\diamond)$ is also a saddle point satisfying 
\begin{equation}
\trace(C^{\diamond\dagger} \alpha C^\diamond) \leq \trace(C^{\diamond\dagger} \alpha^\diamond C^\diamond) \leq \trace(C^{\diamond\dagger} \alpha C^\diamond),
\end{equation} 
for all $(h,\vh,\frakh,C)$ satisfying $\beta = 0$ and $\trace(C^\dagger C) = 1$. Strong duality~\cite{boyd2004convex} implies the optimal value of 
\begin{equation}
\max_{\tilde C}~ \frakF(C^\diamond,\tilde C),\,\text{~~subject~to~}\trace(C^\dagger C)= 1,\,\trace(\tilde{C}) = 0\text{~and~}\trace(J^\dagger_i J_j \tilde{C}) = 0,\forall i,j \in \inull,  
\end{equation}
is equal to that of $\min_{h,\vh,\frakh|\beta = 0}~  4 \trace(C^{\diamond\dagger} \alpha C^\diamond) = \min_{h,\vh,\frakh|\beta = 0}\norm{\alpha}$, proving the optimality of $(C^\diamond,\tilde C^\diamond)$.

\section{\label{app:biased}Highly-biased noise}

We derived the optimal $\frakF$ and the corresponding optimal AQEC code under the highly-biased noise model, taking the limit $\eta \rightarrow 0$. Using the highly-biased model (\eqref{eq:biased}), we need to replace $\vL$ by $(\Pi_\is + \Pi_\iw \sqrt{\eta})\vL$ in \eqref{eq:upper}, where $\Pi_{\iw(\text{or~}\is)}$ is an $r$-by-$r$ diagonal matrix whose $i$-th diagonal element is one when $i\in\iw(\text{or~}\is)$ and zero when $i\in\is(\text{or~}\iw)$. After the following parameter transformation 
\begin{equation}
\vh \leftarrow (\Pi_\is + \Pi_\iw/\sqrt{\eta})\vh,\quad \frakh \leftarrow (\Pi_\is + \Pi_\iw/\sqrt{\eta})\frakh(\Pi_\is + \Pi_\iw/\sqrt{\eta}),
\end{equation} 
in \eqref{eq:upper}, we have the optimal QFI is equal to $\frakF^\diamond = \min_{h,\vh,\frakh|\beta = 0}\norm{\alpha}$, where 
\begin{align}
\alpha &= (\vh\id + \frakh\vL)^\dagger (\Pi_\is + \Pi_\iw/\eta) (\vh\id + \frakh \vL),
\\
\beta &= H + h\id + \vh^\dagger \vL + \vL^\dagger \vh + \vL^\dagger \frakh \vL.
\end{align}
Letting $\balpha = (\vh\id + \frakh\vL)^\dagger \Pi_\iw (\vh\id + \frakh \vL)$, we have
\begin{equation}
\frakF^\diamond =  \frac{1}{\eta}\min_{h,\vh,\frakh|\beta = 0}\norm{\bar{\alpha}} + O(1),
\end{equation}
where $\min_{h,\vh,\frakh|\beta = 0}\norm{\bar{\alpha}} > 0$ as long as $H \notin {\rm span}\{\id,L_i,L_i^\dagger,L_i^\dagger L_j,\,i,j\in\is\}$. 

Now consider the dual program of the modified version of \eqref{eq:reach} with $\alpha$ replaced by $\balpha$. We first simplify the calculation by performing a gauge transformation such that the new set of Lindblad operators $\vJ$ satisfies $\trace(C^\dagger J_i C)=0$, $\scrJ_{\is\is}$ is diagonal with the $i$-th diagonal element equal to $\lambda_i$ when $i\in\is$ and zero when $i\in \iw$, and its Schur complement $\scrJ_{\iw\iw} - \scrJ_{\iw\is}\scrJ_{\is\is}^{-1}\scrJ_{\is\iw}$ is diagonal with the $i$-th diagonal element equal to $\lambda_i$ when $i\in\iw$ and zero when $i\in \is$. Here $^{-1}$ means the Moore-Penrose pseudoinverse and we use the notations $(\cdot)_{\square\blacksquare} = \Pi_{\square} (\cdot) \Pi_{\blacksquare}$ for $\square,\blacksquare = \iw,\is$ and $\scrJ_{ij} = \trace(C^\dagger J_i^\dagger J_j C)$. Note that the gauge transfromation here is divided into two steps (1) $L_i \leftarrow L_i - \trace(C^\dagger L_i C) \id$ and (2) $\vL \leftarrow (u\Pi_\is + v\Pi_\iw) \vL$ where $u = \Pi_\is u \Pi_\is$ and $v = \Pi_\iw v\Pi_\iw$ are some unitary operators within the subspaces defined by $\Pi_{\is}$ and $\Pi_{\iw}$. In this way, the solution is invariant. Again, we introduce a Hermitian matrix $\tilde C$ as a Lagrange multiplier, the Lagrange function is 
\begin{equation}
L(\tilde C,h,\vh,\frakh) = 4\trace(C^\dagger (\vh\id + \frakh\vJ)^\dagger \Pi_\iw (\vh\id + \frakh \vJ) C) + \trace(\tilde C (H + h \id + \vJ^\dagger \vh + \vh^\dagger \vJ + \vJ^\dagger \frakh \vJ)).
\end{equation}
Then we have 
\begin{equation}
\begin{split}
\min_{h,\vh,\frakh|\beta = 0}\norm{\bar{\alpha}} &= \max_{\tilde C} \min_{h,\vh,\frakh} L(\tilde C,h,\vh,\frakh) \\
&= \max_{\tilde C:\trace(\tilde C) = 0} \min_{\vh,\frakh} 4 \big(\vh^\dagger \Pi_\iw \vh + \trace(\frakh^\dagger \Pi_\iw \frakh \scrJ^T )\big) + \trace(\tilde C H) (1 + \vh^\dagger  \vj + \vj^\dagger  \vh + \trace(\frakh \frakj^T))\\
&= \max_{\substack{\tilde C:\trace(\tilde C) = 0,\;\trace(\tilde C H) \neq 0,\\\Pi_\is \vj = 0,\;\frakj_{\is\is} = 0}} \Big( -\frac{1}{4} \trace(\tilde C H)^2 \vj^\dagger \Pi_\iw \vj + \trace(\tilde C H) + (*)\Big),
\end{split}
\end{equation}
and
\begin{equation}
\begin{split}
(*) &=  \min_{\frakh_{\iw\iw},\frakh_{\is\iw}} 4\trace(\frakh_{\iw\iw}\scrJ_{\iw\iw}\frakh_{\iw\iw} + \frakh_{\iw\iw} \scrJ_{\iw\is} \frakh_{\is\iw} + \frakh_{\iw\is} \scrJ_{\is\iw} \frakh_{\iw\iw} + \frakh_{\iw\is} \scrJ_{\is\is} \frakh_{\is\iw}) + \trace(\tilde C H) \trace(\frakh_{\iw\iw} \frakj_{\iw\iw} + \frakh_{\iw\is} \frakj_{\is\iw} + \frakh_{\is\iw} \frakj_{\iw\is}) \\
&= \min_{\frakh_{\iw\iw}} 4\trace(\frakh_{\iw\iw}(\scrJ_{\iw\iw} - \scrJ_{\iw\is}\scrJ_{\is\is}^{-1}\scrJ_{\is\iw})\frakh_{\iw\iw}) + 
\\
&\qquad\qquad  \trace(\tilde C H)  \trace(\frakh_{\iw\iw}(\frakj_{\iw\iw} - \scrJ_{\iw\is}\scrJ_{\is\is}^{-1} \frakj_{\is\iw} - \frakj_{\iw\is}\scrJ_{\is\is}^{-1} \scrJ_{\is\iw}))  - \frac{ \trace(\tilde C H) ^2}{4}\trace(\frakj_{\iw\is}\scrJ_{\is\is}^{-1}\frakj_{\is\iw})\\
&= - \frac{\trace(\tilde C H)^2}{4} \bigg( \sum_{\substack{i,i'\in\bar{\mathfrak{s}},\\\lambda_i+\lambda_{i'}\neq 0}} \frac{\abs{(\frakj_{\iw\iw} - \scrJ_{\iw\is}\scrJ_{\is\is}^{-1} \frakj_{\is\iw} -  \frakj_{\iw\is}\scrJ_{\is\is}^{-1} \scrJ_{\is\iw})_{ii'}}^2}{2(\lambda_i + \lambda_{i'})} + \sum_{\substack{i\in\iw,j\in\is,\\\lambda_j \neq 0}}\frac{\abs{\frakj_{ij}}^2}{\lambda_j}
\bigg),
\end{split}
\end{equation}
under the constraints that $\Pi_{\inull_{\is}}\frakj_{\is\iw} = 0$, $\Pi_{\inull_{\is}} \scrJ_{\is\iw} = 0$, 
\begin{equation}
\Pi_{\inull_{\iw}}\big(\frakj_{\iw\iw} - \scrJ_{\iw\is}\scrJ_{\is\is}^{-1} \frakj_{\is\iw} - \frakj_{\iw\is}\scrJ_{\is\is}^{-1} \scrJ_{\is\iw}\big)\Pi_{\inull_{\iw}} = 0,\label{eq:cond}
\end{equation}
where $\Pi_{\inull_{\is},\inull_{\iw}}$ are projection operators defined by $\inull_{\is} = \{i\in\is|\lambda_i = 0\}$, $\inull_{\iw} = \{i\in\iw|\lambda_i = 0\}$. Otherwise, $(*) = -\infty$.  Note that the second constraint $\Pi_{\inull_{\is}} \scrJ_{\is\iw} = 0$ is automatically satisfied by definition. 

To conclude, the dual program after replacing $\alpha$ by $\balpha$ is equal to 
\begin{equation}
\begin{split}
\max_{C,\tilde C}~ \overline{\frakF}(C,\tilde C),&\text{~~subject~to~}\trace(C^\dagger C)= 1,\,\trace(\tilde{C}) = 0,\;\trace(\tilde{C} H) \neq 0,\\
&\,\,\Pi_\is \vj = 0,\,\frakj_{\is\is} = 0,\,\Pi_{\inull_{\is}}\frakj_{\is\iw} = 0,\, \Pi_{\inull_{\iw}}\big(\frakj_{\iw\iw} - \scrJ_{\iw\is}\scrJ_{\is\is}^{-1} \frakj_{\is\iw} - \frakj_{\iw\is}\scrJ_{\is\is}^{-1} \scrJ_{\is\iw}\big)\Pi_{\inull_{\iw}} = 0.
\end{split}
\end{equation}
where 
\begin{equation}
\overline{\frakF}(C,\tilde C) = \Bigg(\vj^\dagger \Pi_{\iw} \vj + \sum_{\substack{i,i'\in\bar{\mathfrak{s}},\\\lambda_i+\lambda_{i'}\neq 0}} \frac{\abs{(\frakj_{\iw\iw} - \scrJ_{\iw\is}\scrJ_{\is\is}^{-1} \frakj_{\is\iw} - \frakj_{\iw\is}\scrJ_{\is\is}^{-1} \scrJ_{\is\iw})_{ii'}}^2}{2(\lambda_i + \lambda_{i'})} + \sum_{\substack{i\in\iw,j\in\is,\\\lambda_j \neq 0}}\frac{\abs{\frakj_{ij}}^2}{\lambda_j} \Bigg)^{-1}.
\end{equation}

\section{\label{app:dephasing}The optimal AQEC code for correlated-dephasing noise}

In this appendix, we provide a non-perturbation QEC code achieving the optimal $\frakF$ in a correlated noise-dephasing noise model~\cite{layden2018spatial,layden2019ancilla}. We have $N\geq 3$ qubits evolving under 
\begin{equation}
\begin{split}
\frac{d\rho}{dt} &= -i\omega[\vw\cdot\gv{Z},\rho] + \frac{1}{2}\sum_{jk} \Big(\Gamma_{jk} Z_j \rho Z_k - \frac{1}{2}\{Z_kZ_j,\rho\}\Big)\\
&= -i\omega[\vw\cdot\gv{Z},\rho] + \sum_{j} \frac{\mu_j }{2}\Big((\vv_j \cdot \gv{Z}) \rho (\vv_j \cdot \gv{Z}) - \frac{1}{2}\big\{(\vv_j \cdot \gv{Z})^2,\rho\big\}\Big),
\end{split}
\end{equation}
where $Z_i$ is the Pauli-Z operator on the $i$-th qubit, $\vw$ and $\vv_i$ are all unit vertors and $\Gamma = \sum_{i}\mu_i \vv_i \vv_i^T$ with $\mu_i > 0$ and $\{v_i\}$ an orthonormal set of vectors. The HLS condition $H \in \mS$ is equivalent to $\vw \in {\rm span}\{\vv_i,\forall i\}$. 

We first calculate the optimal normalized QFI 
\begin{equation}
4\min_{h,\vh,\frakh|\beta = 0}\norm{\alpha} = 4 \vh^\dagger \vh 
= 2\sum_{i} \frac{(\gv v_i \cdot \gv w)^2}{\mu_i} = 2\vw^T \Gamma^{-1} \vw,
\end{equation}
where $^{-1}$ means the Moore-Penrose pseudoinverse, and
\begin{equation}
\begin{split}
\beta &= H + h\id + \vh^\dagger \vL + \vL^\dagger \vh + \vL^\dagger \frakh \vL \\
& = \vw \cdot \gv{Z} + 
\begin{pmatrix}
\id & \sqrt{\frac{\mu_1}{2}}\vv_{1}\cdot \gv{Z} & \cdots & \sqrt{\frac{\mu_N}{2}}\vv_{N}\cdot \gv{Z}
\end{pmatrix}
\begin{pmatrix}
h & \vh^\dagger \\
\vh & \frakh
\end{pmatrix}
\begin{pmatrix}
\id \\ \sqrt{\frac{\mu_1}{2}}\vv_{1}\cdot \gv{Z} \\ \vdots \\ \sqrt{\frac{\mu_N}{2}}\vv_{N}\cdot \gv{Z}
\end{pmatrix}\\
&= \vw \cdot \gv Z + \sum_{i} (\vh_i+\vh_i^\dagger) \sqrt{\frac{\mu_i}{2}}\vv_{i}\cdot \gv{Z} = 0\qquad \Rightarrow \qquad \vh_i = \frac{\gv v_i \cdot \gv w}{\sqrt{2\mu_i}},\;\frakh = 0.
\end{split}
\end{equation}

Now we introduce a QEC code 
\begin{equation}
\label{eq:phase_code}
\ket{0_\textsc{l}}  =  \bigotimes_{j=1}^N
 \Big(  
\cos\theta_j  \ket{0_j}
 + 
i \sin\theta_j  \ket{1_j}
 \Big),\quad 
\ket{1_\textsc{l}}  =  X^{\otimes N}  \ket{0_\textsc{l}},
\end{equation}
where $\gv\theta = \frac{1}{2}\arccos \chi\vu$, defined element-wise, satisfying
\begin{equation}
    P(\gv v \cdot \gv Z) P = \gv v \cdot \cos(2 \gv \theta) \, Z_\textsc{l}
    =
    \chi (\gv v \cdot \vu )  Z_\textsc{l}, 
\end{equation}
and $P (\gv v \cdot \gv Z) (\gv v' \cdot \gv Z) P \propto P$ for any $\gv v$ and $\gv v'$. $\chi$ is a tunable parameter $\in(0,\norm{\vu}_\infty^{-1}]$ where $\norm{\cdot}_\infty$ is the infinity norm.

The QEC code is designed to correct every mode $\vv$ perpendicular to $\vu$. Using the recovery channel introduced in Appx.~F of \cite{layden2019ancilla}, we would have an effective channel
\begin{equation}
\frac{d\rho}{dt} = -i[\omega \chi (\vu^T \vw) Z_{\tl},\rho] + \frac{(\vu^T \Gamma \vu) \chi^2}{2} \left(Z_{\tl} \rho Z_{\tl} - \rho\right).
\end{equation}
Using spin-squeezed states as input states, we could achieve the optimal $\frakF$ because 
\[
\frakF = \frac{4\chi^2(\vu^T \vw)^2}{2\chi^2(\vu^T \Gamma \vu)} \leq 2 \vw \Gamma^{-1} \vw,
\]
where the second equality holds when 
\[
(\vu^T \Gamma \vu)(\vw^T \Gamma^{-1} \vw) =  (\vu^T \vw)^2 
~~~
\Leftrightarrow~~~  \vu \propto \Gamma^{-1}\vw. 
\]
Note that $\frakF = 2 \vw \Gamma^{-1} \vw$ could be very large if there exist an $i$ such that $\mu_i \ll (\vv_i \cdot \vw)^2$.

\end{document}